# Evolutionary Multi-Objective Aerodynamic Design Optimization Using CFD Simulation Incorporating Deep Neural Network

Yukito Tsunoda [1]
*Fujitsu Limited, Kawasaki, Kanagawa, 211-8588, Japan*

*University of Tokyo, Bunkyo-Ku, Tokyo, 113-8656, Japan*

Akira Oyama [2]

*Institute of Space and Astronautical Science, JAXA, Sagamihara, Kanagawa, 252-5210, Japan*

An evolutionary multi-objective aerodynamic design optimization method using the computational fluid dynamics (CFD) simulations incorporating deep neural network (DNN) to reduce the required computational time is proposed. In this approach, the DNN infers the flow field from the grid data of a design and the CFD simulation starts from the inferred flow field to obtain the steady-state flow field with a smaller number of time integration steps. To show the effectiveness of the proposed method, a multi-objective aerodynamic airfoil design optimization is demonstrated. The results indicate that the computational time for design optimization is suppressed to 57.9% under 96 cores processor conditions.

## I. Nomenclature

| | |
|---|---|
| $AoA$ | = angle of attack |
| $M$ | = Mach number |
| $C_L$ | = lift coefficient |
| $C_D$ | = drag coefficient |
| $C_p$ | = pressure coefficient |
| $e$ | = total energy nondimensionalized by density and sound speed of the ambient condition |
| $Re$ | = Reynolds number based on chord length |
| $r_{LE}$ | = leading-edge radius |
| $x$ | = horizontal coordinate |
| $y$ | = vertical coordinate |
| $L$ | = distance from the surface of the airfoil |
| $u$ | = velocity in *x*-direction nondimensionalized by the speed of sound of ambient conditions |
| $v$ | = velocity in *y*-direction nondimensionalized by the speed of sound of ambient conditions |
| $X_{LO}$ | = lower crest abscissa |
| $X_{UP}$ | = upper crest abscissa |
| $Z_{LO}$ | = lower crest ordinate |
| $Z_{TE}$ | = trailing edge ordinate |
| $Z_{UP}$ | = upper crest ordinate |
| $Z_{XXLO}$ | = lower crest curvature |

---
[1] Senior Researcher, Fujitsu Limited; tsunoda.yukito@fujitsu.com
[2] Associate Professor, Department of Space Flight Systems, 3-1-1 Yoshinodai, AIAA Senior member.

| | |
|---|---|
| $Z_{XXUP}$ | = upper crest curvature |
| $\alpha_{TE}$ | = trailing edge direction |
| $\beta_{TE}$ | = trailing edge wedge angle |
| $\Delta Z_{TE}$ | = trailing edge thickness |
| $\rho$ | = density nondimensionalized by the density of ambient |

## II. Introduction

Multi-objective evolutionary algorithms (MOEAs) have been applied in various fields of aerospace engineering because MOEAs have excellent features (e.g., the capability of identifying Pareto-optimal solutions of a multi-objective design optimization problem in a single run). For example, MOEAs have been applied to optimize spacecraft trajectory design [1,2], earth observation satellite mission planning [3], and rocket engine design [4–7]. They have been applied to aerodynamic design problems such as airfoil shape designs [8–17]. However, an issue is that a large number of design candidates created by MOEAs require to be evaluated using computational fluid dynamics (CFD) simulation, which requires a large computational time [12]. For example, in the flame deflector optimization problem [18], the computational time required to evaluate each design candidate by CFD simulation was up to 7 h with 130 processors (1040 cores) of K supercomputer, and 2,500 design candidates were evaluated. Consequently, this problem took more than two weeks to solve using K supercomputer. Therefore, a problem is that aerodynamics design optimization using an MOEA can only be realized under abundant computing resources. Many approaches have been proposed to address this problem [10, 16, 17, 20, 21, 22].

One method for reducing the computational time is replacing a surrogate model from the CFD simulation [10, 16, 17]. The surrogate model comprises a computationally cheap function such as a neural network [10, 19]. In these studies, first, a certain number of CFD simulations were performed to acquire the sampling data, and a surrogate model was constructed based on these sampling points. Then, the model was used to evaluate the performance of design candidates. In this method, CFD simulations are only needed to construct the model. However, issues remain in terms of the use of surrogate models for aerodynamic design optimization. The optimization of general aerodynamic design problems requires a large number of design parameters and thus high dimensional design space [20, 22]. The number of sample data required to create an accurate surrogate model exponentially grows depending on the increase in the number of design parameters. In the case of a lack of the number of these sample data, inaccuracy caused by the surrogate model is an issue [20, 21]. Moreover, in a high-dimensional design space, the method to select the sampling points is an important problem. The inaccuracy also tends to be more significant in case the model infers a value outside the distribution of the sample data set [23, 24].

Recently, there is a technique to incorporate a deep neural network (DNN) with CFD simulations [25, 26]. In this approach, the DNN infers the flow field, and the CFD simulation starts from the inferred flow field to obtain the steady-state flow field. Using this method, the number of time steps required to reach the steady-state flow was reduced. While the accuracy of the result was not compromised because the performance of the design is evaluated by CFD simulations. This method is suitable for use in the CFD simulation component of MOEAs, wherein various design candidates must be evaluated. However, the effectiveness of this method for multi-objective aerodynamic design optimization has not been clarified. Especially, in the case of using the DNN, the process to prepare the training data and to train the DNN is indispensable. However, there is no report to discuss the total computational cost including this process of training the DNN. So, the MOEA method for implementing this process needs to be developed. In addition, the inference accuracy of the DNN depends on the training data used to train the DNN. Therefore, the method to gather the training data is a critical issue.

The purpose of this study is to propose an evolutionary multi-objective aerodynamic design optimization method using the CFD simulation incorporating the DNN for the steady-state simulation. To realize this, the MOEA method implementing the process of training the DNN is developed. To eliminate the increment of the time of the process to prepare the training data, the grid data and the flow field data, obtained during the process of the MOEA, are used as the training data. The design candidates in the 1st and 2nd generations of MOEA are evaluated using conventional CFD simulations and the DNN is trained using these results. In addition, the obtained results using this method of CFD simulation incorporating the DNN are equal to that of the conventional method. Therefore, these results are also able to use for training data. The method that the DNN is retrained using these results during the MOEA is applied in order to improve the inference accuracy further. In addition, to suppress the increment of time caused by training the DNN, the MOEA method that these processes of training the DNN are parallelly performed with the process of evolution is applied. This MOEA method is evaluated against the sample problem of the design optimization against the 2D airfoil shape. The advantage that the computational time can be suppressed using this method is shown first.

Then the other advantage that the obtained result of this proposed design optimization method is equivalent to that of a conventional method is shown.

## III. Proposed Approach

To reduce the computational time of an aerodynamic design optimization using an MOEA, the CFD simulation method incorporating a DNN is proposed to be used. In this chapter, first, the MOEA method and aerodynamic design optimization using MOEA is described. Then, the CFD simulation method incorporating a DNN is described. After that, the overall flow of a proposed MOEA for this CFD simulation method using the DNN is presented.

### A. Evolutionary Aerodynamic Design Optimization Method

The MOEAs are the optimization method based on the mechanism of evolution. Figure 1 shows the procedures of design optimizations. The parameter sets indicating the designs are generated as the initial populations and evaluated the performance of the parameter sets. The performance values obtained by the evaluation are treated as fitness. The populations of the next generation are reproduced by selection and pairing based on this fitness. By repeating this procedure, design optimization progresses like an evolution. With this procedure, Pareto solutions and optimal design can be obtained automatically.

In the case of the evolutionary aerodynamic design optimization, the evaluations of the fitness are performed by the CFD simulation. Figure 1 shows an example of the optimization of the design of the 2-dimensional airfoil shape. The airfoil shapes are generated depending on the parameter sets. The grid used for CFD simulation is created depending on each airfoil shape, and the CFD simulation is performed to derive the steady-state flow field for each airfoil. The performances of each airfoil shape, such as the lift coefficient ($C_L$) and drag coefficient ($C_D$), are calculated from the derived flow field and these values are supplied as the fitness depending on the parameter sets.

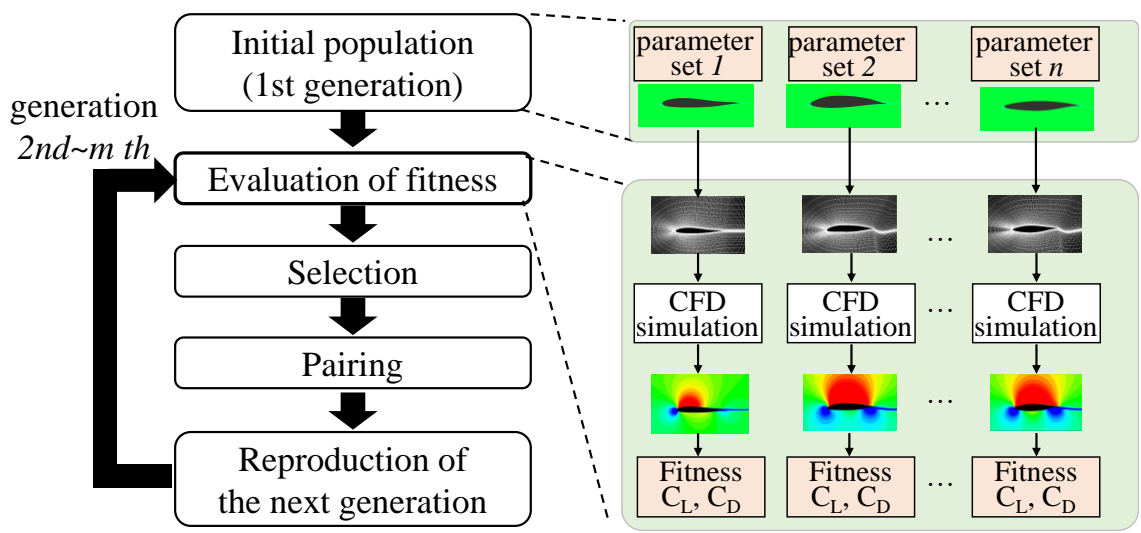

**Fig. 1 Flowchart of evolutionary multi-objective aerodynamic design optimization using CFD simulation.**

### B. CFD Simulation Incorporating a DNN for a Steady-State Flow Simulation

Figure 2 shows the CFD simulation method used to evaluate the performances of designs. As shown in Fig. 2(a), the conventional method for steady-state flow simulation is often used a uniform flow field as an initial state, and the CFD simulation is performed until reaching a steady state. In this method, there is a large difference between the initial state and the steady state. Therefore, the time steps of the CFD simulation of this case are large. To overcome this issue, the method using the DNN inference was reported [25, 26]. The procedure of this method is shown in Fig. 2(b). In the reported method, the CFD simulation used an inferred flow field by the DNN as this initial state. The simulation started from the inferred flow field close to the steady state. Therefore, the time steps can be reduced. Because the computational time of DNN inference is considerably smaller than that of CFD simulation, the total computational time can be reduced.

The DNN uses the position information of the structured grid for the CFD simulation as input information. It can infer the flow field regardless of the position of the grid point, although the position of grid points varies depending on the shape of the design candidates [25, 26]. Therefore, this DNN is suitable for use in the CFD simulation part, which is used to evaluate the various shapes of the design candidates of the MOEA. The output information of the DNN is the values of the flow variables on the grid point, comprising the physical or conserved quantity, such as velocity "$v$" or momentum "$\rho v$". These flow variables can be used as the flow field data of CFD simulation. Thus, the proposed DNN is suitable for inferring the flow field to be applied to the CFD simulation.

In addition, the steady-state flow field is calculated by CFD simulation. Therefore, the flow field obtained is equivalent to that obtained using the conventional CFD simulation, even if the accuracy of DNN inference degrades. As a result, the calculated performance values of $C_L$, and $C_D$ are accurate. And then, the result of the proposed design optimization method is also equivalent to that of a conventional method.

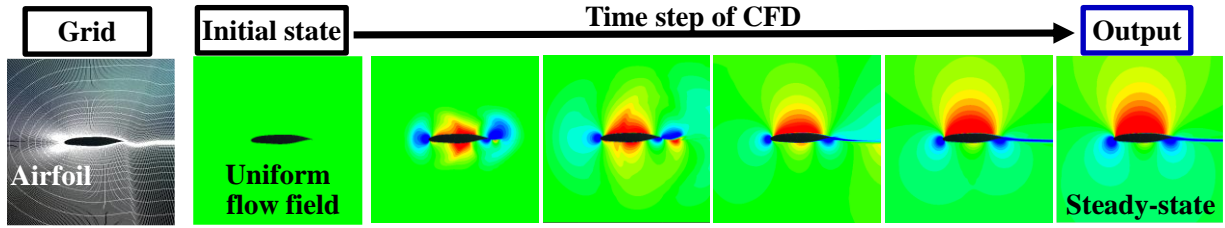

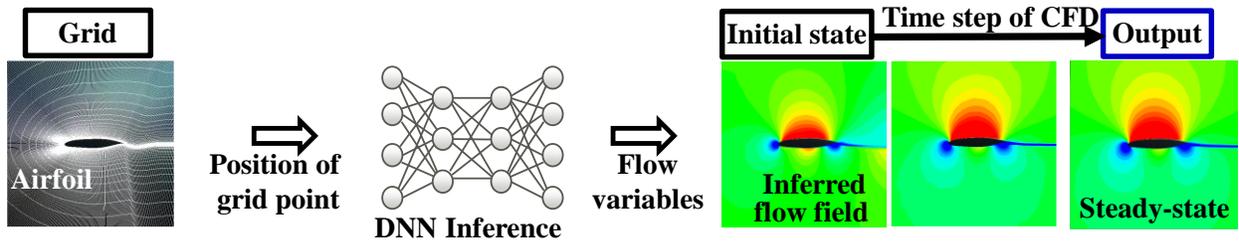

**Fig. 2 CFD simulation of a conventional CFD simulation (a) and CFD simulation incorporating the DNN (b). Source: Adapted from [26]. In CFD simulation incorporating the DNN, a steady-state flow field is inferred by the DNN and CFD simulation is performed from this inferred flow field.**

### C. Evolutionary Aerodynamic Design Optimization Method Including the Process of Training the DNN

To use this CFD simulation using the DNN, the process of preparing the training data and training the DNN is required. Therefore, the MOEA including the process of training the DNN is developed this time. The procedure of this MOEA is shown in Fig. 3. The designs of the 1st and 2nd generations in the MOEA are evaluated using the conventional CFD simulation. Then, the DNN is trained using the grid and flow field data obtained on these generations. After the DNN is trained, the CFD simulation part is replaced with a CFD simulation incorporating a DNN.

In the MOEA, new design candidates of the latest generation are often generated outside of the data distribution of the design candidates of a previous generation. These are caused by an important property of the MOEA that enables the generation of an unknown superior design by generating certain design candidates outside of the data distribution of design candidates of previous generations [27]. However, at the initial training stage of the DNN, these design candidates have not yet been generated and are not included in the training dataset. The DNN has a property that the inference accuracy for these design candidates degrades [23, 24]. The time steps of CFD simulation increase when the inference accuracy of the DNN degrades. However, the derived flow field by CFD simulation is accurate regardless of whether a DNN is used. Therefore, the obtained result during the MOEA process can also be used as training data. The inference accuracy can be further improved by training the DNN using the data including these latest generations of the MOEA. Thus, the DNN is retrained using the grid and flow field data including the latest generations. After the DNN is retrained, the DNN part is replaced with the updated one. In other words, because it is assumed that the DNN will be updated during the process of design optimization in this method, the DNN trained with an insufficient number of training data can be applicable at the initial stage.

The increment of the total computational time caused by the process of training the DNN should be suppressed. Therefore, the MOEA method that these processes of training the DNN are parallely performed with the process of evolution is applied. At the initial stage, during the DNN is trained, the MOEA is continued by using the conventional CFD simulation. After the DNN is trained, the CFD simulation part is replaced with a CFD simulation incorporating a DNN. Similarly, the processes of retraining the DNN are parallely performed with the process of evolution. After the process of retraining is completed, the DNN part is replaced with the updated one.

Here, in the case of retraining the DNN, two different training data sets can be created: a dataset comprising of the designs of the latest generation (e.g., design candidates of the 11th and 12th generations), and a data set comprising all designs at that point (e.g., design candidate from the 1st to 12th generations). Therefore, each case is evaluated this time using the method of "*CFD simulation incorporating an evolution update-type DNN using latest designs*" (Method 1) and the method of "*CFD simulation incorporating an evolution update-type DNN using all designs*" (Method 2).

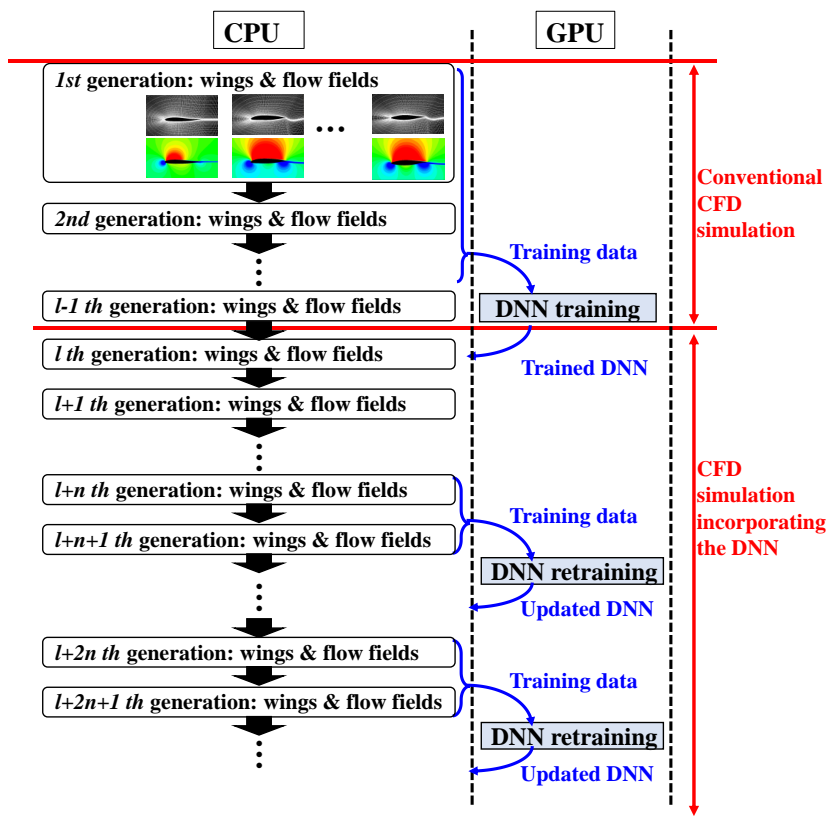

Fig. 3 MOEA method using the CFD simulation incorporating the DNN including the process of gathering the training data and training the DNN.

## IV. Experimental Setup

In this section, first, the sample problem used to confirm the effect of our proposed evolutionary aerodynamic design optimization method is described. Then, the architecture and libraries used for the evaluation are shown.

### A. Sample Problem of an Aerodynamic Design Optimization

The optimization of a 2D airfoil shape in a steady-state flow field is used as the sample problem. The objective is to find the airfoil shapes that have the property of maximized $C_L$ and minimized $C_D$. The evaluation is performed with a Reynolds number of 1,500,000, a Mach of 0.15, and an AoA of 1.1°, based on the reported conditions in [28]. The PARSEC parameters [10, 16, 29] shown in Fig. 4 are used as parameters to create an airfoil shape for design optimization. Table 1 presents the search space of the design optimization's parameters.

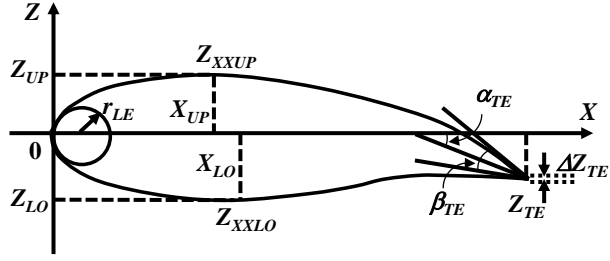

**Fig. 4 PARSEC airfoil parameters.**

**Table 1. Parameter ranges of the design space**

|  | $r_{LE}$ | $X_{UP}$ | $Z_{UP}$ | $Z_{XXUP}$ | $X_{LO}$ | $Z_{LO}$ | $Z_{XXLO}$ | $Z_{TE}$ | $\alpha_{TE}$ (rad) | $\beta_{TE}$ (rad) |
|---|---|---|---|---|---|---|---|---|---|---|
| Minimum, limit | 0.0055 | 0.25 | 0.048 | -1.0294 | 0.25 | -0.071 | -0.0686 | -0.02 | -0.3580 | 0.0201 |
| Maximum limit | 0.0215 | 0.6043 | 0.1194 | -0.418 | 0.5376 | 0.00 | 0.8204 | 0.02 | 0.02304 | 0.2571 |

**B. Computational Method**

The simulator calculates the flow field by solving the compressible Navier–Stokes equations. In this study, LANS3D is employed as the CFD simulator [30]. Upwind-biased 2nd-order differencing [31], the alternative direction implicit-symmetric Gauss-Seidel (ADI-SGS) scheme [32], and the turbulence model of Baldwin and Lomax [33] are used. The structured grid is created following the method described in a previous paper [34]. The flow field is treated to reach a steady state if the variation of the $C_D$ value was below 2e-4 during the 1,000 steps, which corresponds to the error in the $C_D$ value is almost 1 count or below.

A residual neural network (ResNet) is employed as the DNN architecture to infer the flow field [26, 35, 36]. The DNN comprises an encoder and a decoder [26]. The architecture of our DNN is shown in Fig. 5. In the encoder part, the features of the airfoil shape are extracted. In the decoder part, the flow field is inferred based on the extracted features. Both the encoder and decoder comprise four units, and each unit includes convolutional and shortcut connections. This DNN is implemented using Keras [37] and a TensorFlow GPU 1.13 backend.

The MOEA is implemented based on a previous study. The elitist mate selection based on the binary tournament (termed EBT) is used as the mate selection scheme [38]. The MOEA/D-multi-objective to multi-objective (MOEA/D-M2M) decomposition is applied [39]. Simulated binary crossover (SBX) is used as the crossover operator [27]. The MOEA was performed with 96 populations and 100 generations.

The CFD simulation is run on the Intel Xeon Gold 5218 (2.30 GHz) while DNN training is performed on the dual NVIDIA Tesla P100 GPU.

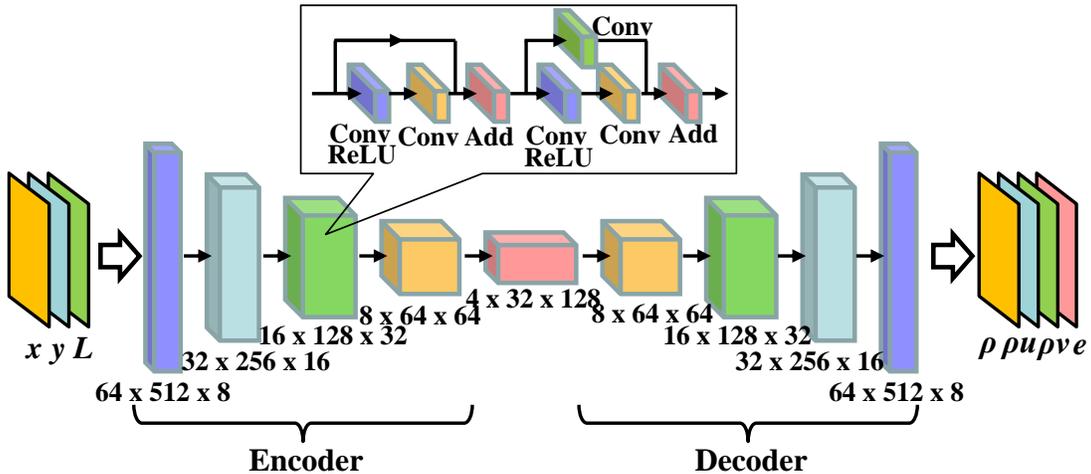

**Fig. 5 The architecture of DNN used to infer the flow field**

# V. Results

The advantage of our proposed method is realized to reduce computational costs while maintaining the quality of the optimization results as the conventional method. Therefore, the design optimization by MOEA was performed both with the conventional method (without DNN) and with the method using the DNN and then the results were compared. In addition, to confirm the effect to update the DNN, the method that the DNN trained with 1st and 2nd generation is used whole MOEA process was also evaluated as the method of "*CFD simulation incorporating a fixed-type DNN*". The property that the computational time can be suppressed was shown by comparing the method using the DNN with the conventional method. Then, the property that the quality of the optimization results was equal was shown by comparing the obtained designs and Pareto solutions of each method.

## A. Evaluation of Total Computational Time of the Evolutionary Aerodynamic Design Optimization
### *1. Execution Time of Each Process*

To clarify the execution procedure of the MOEA method shown in Fig. 3, the computational time of each process was confirmed. The computational time of the CFD simulation is proportional to the number of time steps. Therefore, the computational time of the CFD simulation part was evaluated by using this number of time steps. The CFD simulation execution of the MOEA is performed for each generation. Therefore, the computational time was evaluated for each generation of the MOEA. In case the aerodynamic evaluation was performed with the processors of 96 cores total, all the CFD simulations of one generation are performed parallelly. In this case, the computational time of one generation depends on the design whose computational time is worst in one generation. On the other hand, in case the aerodynamic evaluation was performed with the single-core processor, the CFD simulation is performed serially. In this case, the computational time of one generation depends on the average time of the simulation. Therefore, the time steps of the worst value and the average time steps in each generation were evaluated.

As an example of the computational time of CFD simulation, the computational time of the 30th generation was presented in Table 2. The required number of time steps of each CFD simulation method against the worst design candidate was 14,800 steps with the conventional method, 11,000 steps using the method with fixed-type DNN, 9600 steps using the evolution update-type DNN using the latest design, and 9600 steps using with the evolution update-type DNN using all designs. Each 10,000 CFD simulation step takes 0.92 h. These were equivalent to the required computational time of 1.36, 1.01, 0.88, and 0.88 h respectively. The number of average required time steps of each CFD simulation method against one design candidate was 10,460 steps with the conventional method, 7,058 steps using the method with fixed-type DNN, 5,269 steps using the evolution update-type DNN using the latest design (Method 1), and 5,800 steps using with the evolution update-type DNN using all designs (Method 2). These were equivalent to the required computational time of 0.96, 0.65, 0.48, and 0.53 h respectively. Training the DNN took 8.5 h with the GPU. The DNN should be retrained to update the DNN for the evolution update-type DNN. In this case, the transfer learning technique was applied. Therefore, retraining the DNN with the GPU can be suppressed to 1.6 h. Moreover, inferring the flow field by the trained DNN with a 1-core CPU took less than 1 min. Therefore, the total execution time of the design optimization is regarded as the execution time of CFD simulation and DNN training.

**Table 2 Computational time of each method and each process**

| Method of CFD simulation part | Time for DNN training | CFD time/1 sample · 1 core CPU (Gen.30, average) | CFD time/1 sample · 1 core CPU (Gen.30, worst) |
|---|---|---|---|
| Conventional CFD simulation (Without DNN) | - | 0.96 h | 1.36 h |
| CFD simulation with fixed-type DNN | 8.5 h | 0.65 h | 1.01 h |
| CFD simulation with evolution update-type DNN using the latest designs (Method 1) | 8.5 h (Initial) 1.6 h (update) | 0.48 h | 0.88 h |
| CFD simulation with evolution update-type DNN using all designs (Method 2) | 8.5 h (Initial) 1.6 h (update) | 0.53 h | 0.88 h |

*2. Required Time of the Entire Design Optimization*

The total computational time of the MOEA method shown in Fig. 3 was calculated using these results. In the 1st and 2nd generations whose data were used as the training data, the designs were evaluated with the conventional CFD simulation. And the conventional CFD simulation was used also for the evaluation of the designs in the period during which the DNN has been trained in parallel. The training time of the DNN was 8.5 h with the GPU. This duration was equivalent to that of the 7 generations in the MOEA process under the 96-core CPU condition. Therefore, the 1st and 2nd generations were used to gather the training data and from the 3rd to 9th generation where the DNN had been trained parallelly, the evaluation was performed with the conventional CFD simulation. And from the 10th to the last generation, the evaluation was performed with the CFD simulation incorporating the DNN.

To reduce the computational time further by improving the inference accuracy of the DNN, the DNN was retrained and updated as shown in Fig. 3. The retraining of the DNN was performed every 10 generations in this research. The computational time to retrain the DNN was 1.6 h. This duration was equivalent to that of the 2 generations in the MOEA process under the 96-core CPU condition. During the retraining of the DNN which is equivalent to 2 generations in the MOEA process, the CFD simulation was performed with the DNN before the update. And after the DNN was retrained, the CFD simulation was performed using the updated DNN. On the other hand, with the method of "*CFD simulation incorporating a fixed-type DNN*", the DNN trained using the results of 1st and 2nd generations was used during the whole the 10th to the last generation.

To evaluate the effect when the computational time is reduced, the result of the method utilizing conventional CFD simulation was used as a reference. Figure 6 shows the results of the number of time steps depending on the generation of the MOEA. Figures 6(a) and 6(b) show the worst number of steps of the design of each generation and the average number of steps of each generation, respectively. The horizontal axis shows the generation number of the MOEA, and the vertical axis shows the average or worst number of time steps required for each generation, respectively. Because the CFD simulation incorporating the DNN was started to be used from the 10th generation, the required number of time steps was reduced from the 10th generation. The DNN was updated every 10 generations. On this timing, a further reduction of the required time steps was observed. This was likely caused by the improvement of the inference accuracy by retraining the DNN including the latest designs. The number of time steps was further reduced when the method of "*CFD simulation incorporated the evolution update-type DNN using the latest designs*" (Method 1) compared to "*using all the designs*" (Method 2). The reason was likely that the DNN was specialized to the optimal designs created by the MOEA and the inference accuracy was improved against these designs.

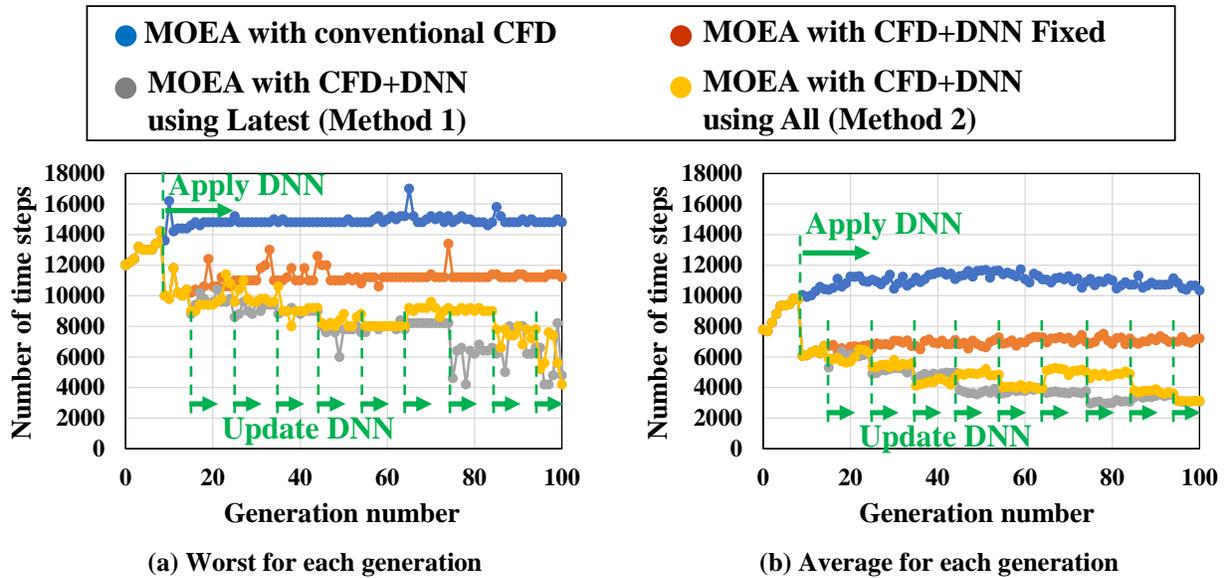

(a) Worst for each generation  (b) Average for each generation

**Fig. 6 The number of time steps of CFD simulation depending on the generation of MOEA.**

First, the total execution time of design optimization performed with a 96-core CPU was evaluated. In this case, the evaluation of the design candidates using CFD simulations is performed parallelly, where the execution time of each generation depends on the largest execution time in each generation. The results were shown in Fig. 7(a). The execution times of design optimization using the method of "*CFD simulation incorporating evolution update-type DNN using the latest designs*" were 78.3 h, which is 57.9% of a conventional method. The total execution time using the method of "*CFD simulation incorporates the fixed-type DNN*" was 104.2 h, which was 76.9% of a conventional method. This difference indicated the effect to update the DNN depending on the evolution. The execution times of design optimization using the method of "*CFD simulation incorporating evolution update-type DNN using all designs*" were 84.3 h which is 62.3% of a conventional method. Therefore, the latest designs were suitable to use when updating the DNN with the CFD using the evolution update-type DNN.

Then, we evaluated the total execution time in the case of the design optimization performed with a 1-core CPU. The result was shown in Fig. 7(b). Like the results under using a 96-core CPU, the computational time was suppressed most in the case of using the method of "*CFD simulation incorporating evolution update-type DNN using the latest designs*". The execution times of design optimization using this method was 4152 h, which was 43.5% of the conventional method.

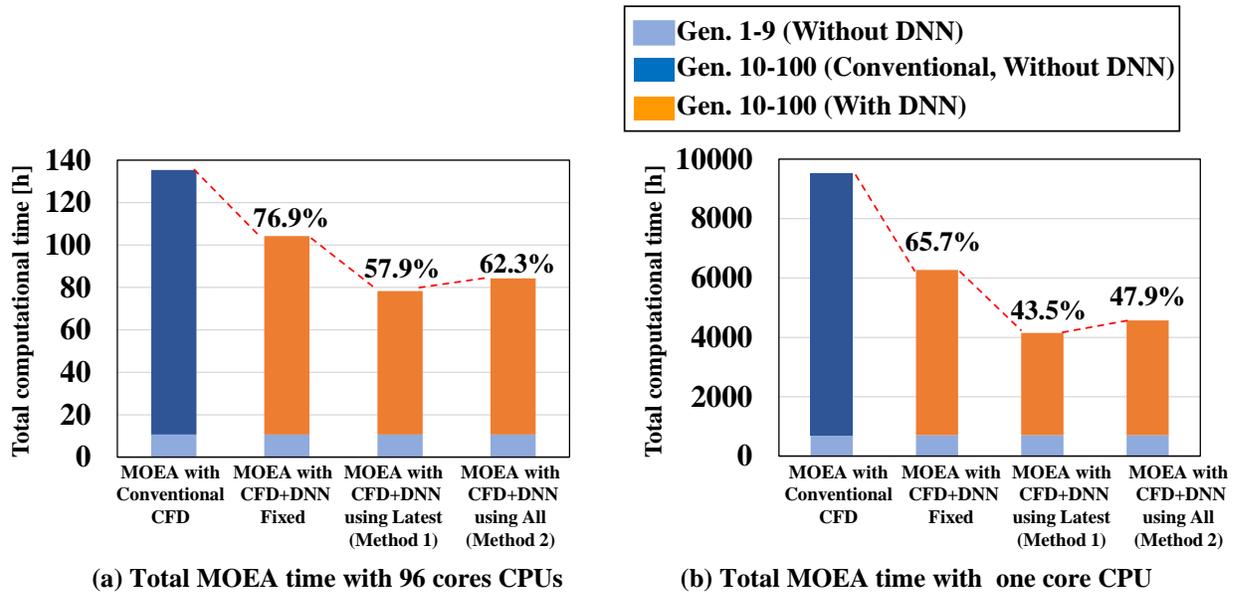

**(a) Total MOEA time with 96 cores CPUs**     **(b) Total MOEA time with one core CPU**

**Fig. 7 Comparison of the results of the total computational time**

**B. Comparison of the Results of the Design Optimization**

The property that the quality of the optimization results of this proposed method was equal was confirmed by comparing the conventional method. As a preliminary, the design optimization using the conventional MOEA method was performed to confirm the design optimization correctly under the evaluated condition. The distribution of design candidates during the process of evolution was evaluated first. Figure 8 showed the distribution of the obtained solutions. Figures 8(a) and (b) showed the distribution of the evaluation results of the design candidates of the 1st and 2nd and 11th and 12th generations, respectively. As shown in these results, superior designs were confirmed to be generated as evolution progresses. In particular, design candidates that had higher $C_L$ values were confirmed to be newly generated by evolution.

All the design candidates and Pareto solutions obtained with 100 generations of the MOEA were shown in Fig. 9(a). The airfoil shape with $C_L$=0.8 on the Pareto front was observed. The airfoil shape and the pressure coefficient fields around it was shown in Fig. 10(a). The obtained airfoil shape had the features of the increased camber, as reported in the previous research [28]. With these results, the optimization was confirmed to be performed properly under this condition.

Each design optimization method using the CFD simulation incorporating the DNN was also performed with the same condition. All the design candidates obtained by each method were compared in Fig. 9(b)-(d). No significant difference in the distribution of Pareto-optimal solutions against that obtained using the conventional method was observed. The airfoil with minimum $C_D$ at $C_L$=0.8 was also indicated and compared with that of the previous research.

The airfoil shape and the pressure coefficient fields around it was shown in Fig. 10(b)-(d). The obtained airfoil shape with all the methods was found to be almost the same, and the features were also the same as that of the reported one in the previous research [28]. With these results, the property that the optimization results with the method using the DNN were equal to that with the conventional method was confirmed.

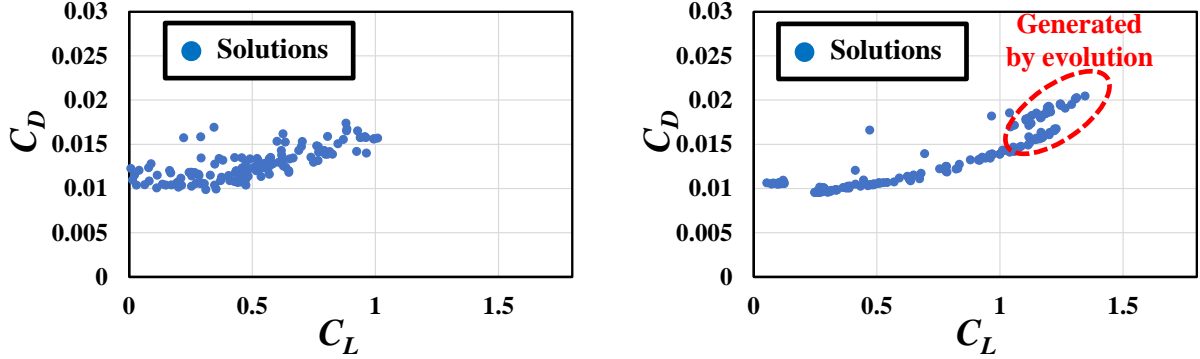

(a) Design candidates of 1st and 2nd generations  (b) Design candidates of 11th and 12th generations

Fig. 8 Distribution of the solutions.

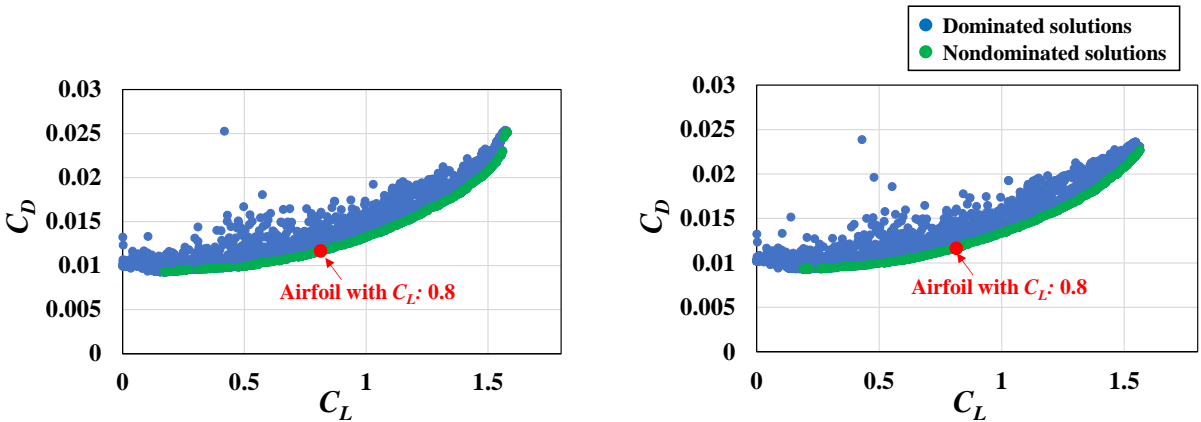

(a) The conventional CFD simulation

(b) CFD simulation incorporating a fixed-type DNN

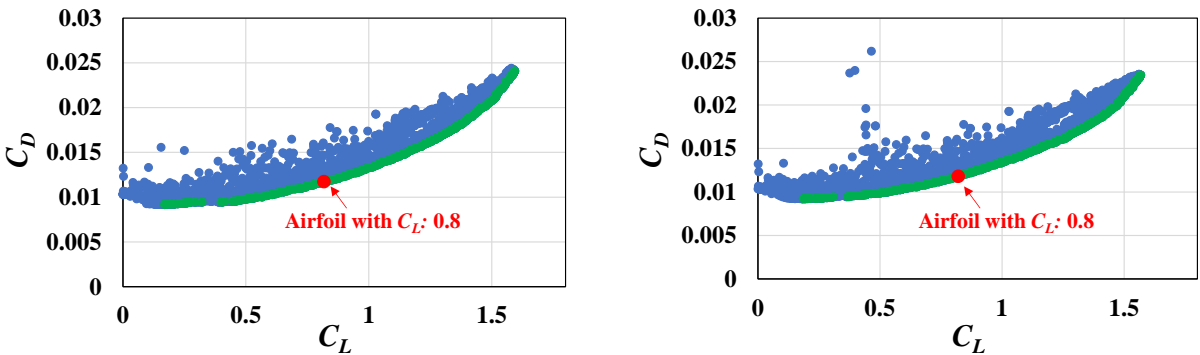

(c) CFD simulation incorporating an evolution evolution update-Type DNN using the latest designs (Method 1)

(d) CFD simulation incorporating an update-type DNN using all designs (Method 2)

**Fig. 9 Results of design optimization. Distribution of the Pareto-optimal solutions and other design candidates.**

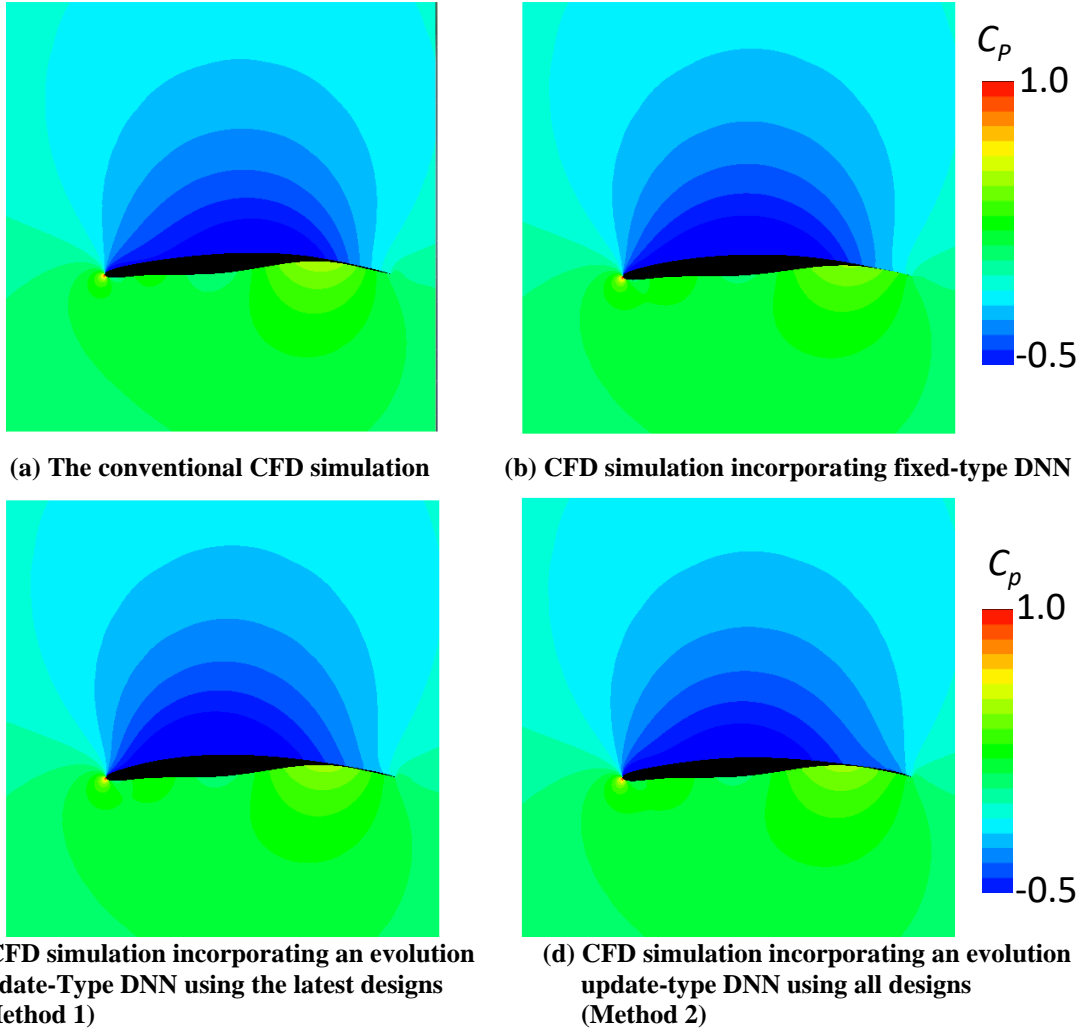

**Fig. 10 The airfoil shapes with $C_L$=0.8 on the Pareto front in Fig. 9. The airfoil shape and the pressure coefficient field around the airfoils were shown.**

## VI. Conclusions

An evolutionary multi-objective aerodynamic design optimization method using the CFD simulation incorporating a DNN to reduce the required computational time was proposed. In this approach, the DNN inferred the flow field from the grid data and the CFD simulation started from the inferred flow field to obtain the steady-state flow field with a smaller number of time integration steps. In the proposed optimization approach, the design candidates in the 1st and 2nd generations were evaluated using the conventional CFD simulation and then the DNN was trained using grid data and flow field data of these design candidates. The DNN was updated also after a certain number of generations using the CFD simulation results of the design candidates of the latest generation. To suppress the increment of time caused by the process of training the DNN, the process of the DNN training was performed parallelly with the process of evolution of the MOEA. After that, the DNN trained with the data started to be used.

To show the effectiveness of the proposed method, a multi-objective aerodynamic airfoil design optimization was demonstrated. The computational time was suppressed to 57.9% when the aerodynamic evaluation using CFD simulation was parallelized using 96 processors. And also, the results showed that the computational time was suppressed to 43.5% if a single processor was used for the optimization.